# Alloharmonics in Burst Intensification by Singularity Emitting Radiation

K. Ogura[1*], M. S. Pirozhkova[1*], A. Sagisaka[1*], T. Zh. Esirkepov[1], A. Ya. Faenov[2†], T. A. Pikuz[2], H. Kotaki[1], Y. Hayashi[1], Y. Fukuda[1], J. K. Koga[1,3], S. V. Bulanov[1,4], H. Daido[5], N. Hasegawa[1], M. Ishino[1], M. Nishikino[1], M. Koike[1], T. Kawachi[1], H. Kiriyama[1], M. Kando[1], D. Neely[6,7†], and A. S. Pirozhkov[1‡]

[1.] Kansai Institute for Photon Science (KPSI), National Institutes for Quantum Science and Technology (QST), 8-1-7 Umemidai, Kizugawa, Kyoto 619-0215, Japan
[2.] Institute for Open and Transdisciplinary Research Initiatives, Osaka University, Suita, Osaka 565-0871, Japan
[3.] Kyoto International University Academy, 63-1 Yuden Tanabe, Kyotanabe Kyoto 610-0331, Japan
[4.] Institute of Physics of the Czech Academy of Sciences, Extreme Light Infrastructure Beamlines Project, Na Slovance 2, 18221 Prague, Czech Republic
[5.] Institute of Laser Engineering, Osaka University, 2-6 Yamadaoka, Suita, 565-0871 Osaka, Japan
[6.] Central Laser Facility, Rutherford Appleton Laboratory, STFC, Chilton, Didcot, Oxon OX11 0QX, United Kingdom
[7.] Department of Physics, SUPA, University of Strathclyde, Glasgow G4 0NG, United Kingdom

**ABSTRACT**. Burst Intensification by Singularity Emitting Radiation (BISER) in underdense relativistic laser plasma is a bright source of coherent extreme ultraviolet (XUV) and x-ray radiation. In contrast to all harmonic generation mechanisms, high-resolution experimental BISER spectra in the XUV region contain spectral fringes with separation much finer (down to 0.12 eV) than the initial driving laser frequency (~1.5 eV). We show that these fringe separations result from two main factors: laser frequency downshift (redshift) due to the quasi-adiabatic energy loss to the plasma waves, and spectral interference of different harmonic orders from different emission moments, i.e. alloharmonics [simultaneously submitted paper: Pirozhkova et al., *Phys. Rev. Research* (2025)].

Catastrophe theory [1] states that multistream flows always produce density singularities. If a medium contains elementary emitters capable of travelling wave emission, these multistream flow singularities lead to coherent emission called Burst Intensification by Singularity Emitting Radiation (BISER) [2]. BISER is a collective effect, proportional to the number of coherent elementary emitters squared; it originates from singularities with dimensions much smaller than the driver characteristic spatial scale, and therefore BISER is much brighter than the singularity-free medium emission. In the case of an underdense plasma driven by a relativistically intense laser with the dimensionless amplitude $a_0 > 1$ driving a bow wave [3], BISER is a bright source of temporally and spatially coherent XUV and x-ray radiation [4,5]; here $a_0 = eE_L/m_e c\omega_L = (I_L/I_R)^{1/2}$ [6], $e$ is the elementary charge, $m_e$ is the electron mass, $I_L$ is the peak irradiance, and $I_R = \pi m_e^2 c^5 / 2e^2 \lambda_L^2 \approx 1.37 \times 10^{18}$ W/cm$^2 \times (\lambda_L[\mu m])^{-2}$ for a linearly polarized laser with angular frequency $\omega_L$ and wavelength $\lambda_L$; $\lambda_L[\mu m]$ denotes the laser wavelength in micrometers. Based on the BISER brightness at $P_L \sim 20$ TW laser power [2] and BISER yield quadratic scaling, $\propto P_L^2$, demonstrated up to 120 TW [4], the brightness of BISER driven by petawatt-class lasers can exceed XUV Free Electron Lasers. BISER produces attosecond pulses with duration close to the transform limit [2], which together with a multi-hundred-eV bandwidth [7] promise pulses shorter than the atomic time unit of 24 attoseconds, the major goal of the attosecond physics.

One feature of BISER, evident already from its discovery in the soft x-ray spectral region [4], is a complex spectral structure containing long-period modulations and various sets of fine fringes (e.g., Figs. 5-7 and 9 in [5]). The long-period modulations have periods from several eV to several tens and even hundreds of eV, i.e. significantly larger than the driving laser photon energy, $\hbar\omega_L \sim 1.5$ eV in the experiments. These modulations, determined by the sub-cycle (attosecond) pulse separations, have been studied in [8]. Here we focus on the fine spectral fringes determined by longer (femtosecond and tens of femtosecond) pulse separations, which are key to the BISER attosecond pulse train properties.

---

[*] These authors contributed equally to this work
[†] Deceased
[‡] Contact author: pirozhkov.alexander @ qst.go.jp

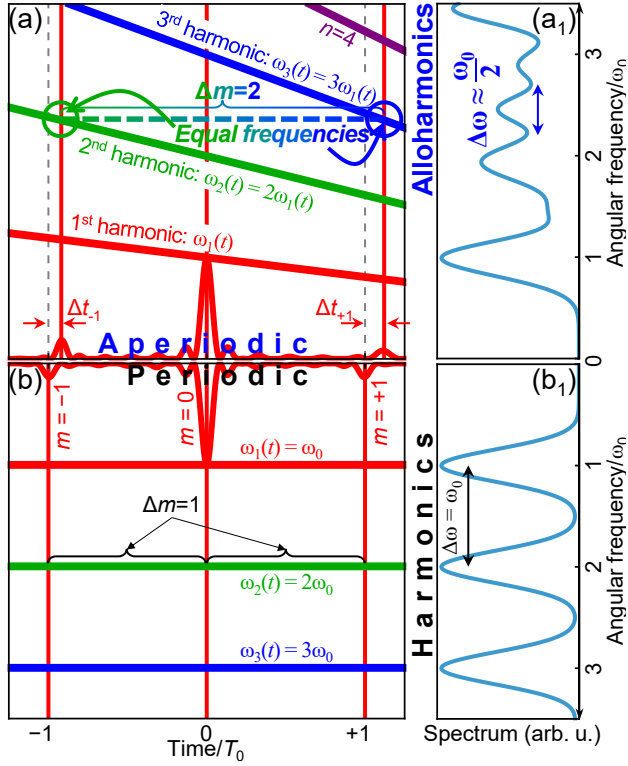

FIG. 1. Formation of alloharmonic (a,a$_1$) and harmonic (b,b$_1$) fringes in spectra of aperiodic (thick red line in a) and periodic (thick red line in b) pulse trains defined by the temporal phase $\Phi(t)=\omega_0 t(1+\alpha(t/T_0))$ with the driver frequency $\omega_0$ and period $T_0=2\pi/\omega_0$ and dimensionless chirp parameter $\alpha$, which gives individual pulses at $t_m=T_0([1+4\alpha m]^{1/2}-1)/2\alpha$ with frequencies $\omega_m=\omega_0(1+4\alpha m)^{1/2}$ ($t_m=mT_0$ and $\omega_m=\omega_0$ for $\alpha=0$) [9]. Red vertical lines in (a,b) show positions of the pulses in time, note the pulse shifts $\Delta t_{-1}$ and $\Delta t_{+1}$ in (a); the colored horizontal lines show dependence of harmonic frequencies vs time $\omega_n=n\omega_0(1+2\alpha(t/T_0))$ with $\alpha=0$ in (b) and $\alpha=-0.096$ in (a); for this $\alpha$ value, the 2$^{nd}$ harmonic of the pulse $m=-1$ has the same frequency as the 3$^{rd}$ harmonic of the pulse $m=+1$, thus producing *alloharmonics* with $\Delta m=2$ (horizontal dashed line), i.e. spectral fringes with separation $\Delta\omega\approx\omega_0(1/\Delta m+\alpha[1+2m/\Delta m])=\omega_0/2$ in (a$_1$) [9].

At first glance, BISER fine spectral fringes resemble laser harmonics. However, all harmonic generation mechanisms have frequency components $(n+\delta n)\omega_L$ with integer values of $n$ and possible constant ($n$-independent) shifts $\delta n$, $0\leq\delta n<1$, caused by the carrier-envelope phase [10]. Due to specific symmetries, some harmonic orders can be absent: thus, atomic harmonics typically contain odd orders [11], while harmonics from solid targets may contain all, even, or odd orders, depending on the polarization [12,13]. Importantly, in all cases the harmonic separations equal integer *multiples* of the laser frequency:

$\Delta\omega=\Delta n\omega_L$, with integer $\Delta n=n_2-n_1$. Surprisingly, BISER exhibits fine fringe separations *smaller* than the driving laser frequency. The explanation to this observation proposed in [4] is the gradual laser frequency redshift caused by the energy depletion as the laser pulse transfers its energy to the wake wave and bow wave, while the number of laser photons is approximately conserved [14,15]. Significant redshifts by this mechanism, with frequency changes of more than a factor of ten, have been observed experimentally [16].

Yet, another explanation of experimentally observed harmonic-like fringes in spectra of radiation reflected from relativistic plasma surfaces [17-19] is the aperiodic pulse train. Recently this mechanism of fine spectral fringe formation, *alloharmonics*, has been explained in detail [20]. Alloharmonic fringes appear in the case of a nonperiodic or an inexactly periodic driver, as a result of spectral interference of *equal* frequencies of *different* harmonic orders from *different* driver cycles, Fig. 1. For example, if the frequency changes by 1% from $\omega_1$ to $\omega_2=0.99\omega_1$ in one cycle, the harmonic number 99 from the first cycle would interfere with harmonic number 100 from the second cycle, since $99\omega_1=100\omega_2$. In general, if (integer) $m$ is the cycle number (where $m=0$ is the central cycle, $m<0$ and $m>0$ correspond to earlier and later cycles), (integer) $\Delta m>0$ is the separation between the two cycles, $\omega_m$ and $\omega_{m+\Delta m}$ are the corresponding angular frequencies, and $n$ and $n+\Delta n$ are the corresponding harmonic orders, the *alloharmonic equation* is:

$$(1) \quad n\,\omega_m = (n+\Delta n)\,\omega_{m+\Delta m}.$$

A generalization to non-zero carrier-envelope phase is provided in [20]. The spectral interference of equal frequencies in (1) produces spectral fringes with the separation $\Delta\omega=2\pi/(t_{m+\Delta m}-t_m)$, where $t_{m+\Delta m}-t_m$ is the temporal separation between the two interfering cycles. Under small-nonperiodicity conditions which are typical for many important cases including our experiments, we can approximate each cycle duration as $T_0$, i.e. the central cycle duration, or driver "period" (we will use the driver period and driver frequency in this sense, namely the central cycle duration and its reverse, without quotes even for nonperiodic drivers). Thus, in small-nonperiodicity cases, $t_{m+\Delta m}-t_m\approx\Delta m T_0$, and $\Delta\omega\approx 2\pi/\Delta m T_0=\omega_0/\Delta m$: the allohamonic fringe separations are *approximately* equal *integer fractions* of the driver frequency, and thus, can be several times smaller than it, even though the driver frequency change causing these may be on the percent level. Alloharmonics with $\Delta m=1$ produce spectral fringes with separations approximately (but *not exactly*) $\omega_0$ and can be misleadingly treated as harmonics, which in this case causes a wrong determination of the driver parameters, in particular, $\omega_0$ and carrier-envelope phase, because alloharmonic positions can be different from $(n+\delta n)\omega_0$, as we will show later.

The question is, which of these two mechanisms, or possibly their combination, is responsible for the experimental observations.

To study the fine BISER spectral fringes in detail, we performed an experiment with laser and plasma parameters as in [4,5], however, we observed BISER in the softer spectral region (17-34 nm), where the spectrograph had much higher resolving power, so much finer fringes could be observed compared to the earlier experiments. This allowed us to resolve fringes with separation down to ~0.12 eV, i.e. approximately 13 times smaller than the laser photon energy. From the accumulated statistics, we found that the observed fine fringes are explained by the combination of the two factors: gradual laser redshift with typical factors ~1.5-3 in our case and alloharmonics providing additional discrete close-to-integer factors.

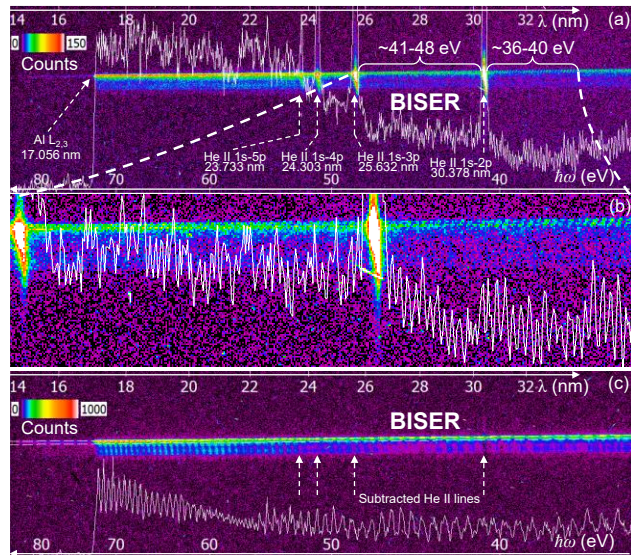

FIG. 2. Experimental BISER spectra. (a) Example with fine spectral fringes, (b) shows their magnified view; the Fourier analysis of this shot is shown in Fig. 3(a). (c) Example with changing fringe spacing; the analysis of this shot is shown in Fig. 4.

The experiment was performed with the linearly polarized J-KAREN-P laser [21]. The laser pulses had a central wavelength $\lambda_L$=0.8115±0.0007 μm, corresponding to the photon energy $\hbar\omega_L$=1.5278±0.0013 eV; here and throughout this paragraph, the errors listed are the standard deviation of the shot-to-shot parameter variations. The pulse energy was $\mathcal{E}_L$=0.648±0.011 J. The pulse duration and spot shape were measured in vacuum at full laser power attenuated with two wedges and additional high-quality neutral-density filters in the case of the spot measurement. The pulse measured with self-referenced spectral interferometry [22] had a full width at half maximum (FWHM) duration of 33±5 fs and an effective duration of $\tau_{Eff}$=38±6 fs. Here, $\tau_{Eff}=\int P_n(t)dt$ is the area under the normalized power curve $P_n(t)$ [23]. The peak power $P_L=\mathcal{E}_L/\tau_{Eff}$ was 17±3 TW. The pulses were focused with an $f/9$ off-axis parabolic mirror. The focal spot, measured with an achromatic lens and high dynamic range charge-coupled device (CCD), had an FWHM of (10.1±0.2) μm × (12.1±0.5) μm, horizontal (along the polarization direction) × vertical, and an effective spot radius $r_{Eff}$=8.3±0.3 μm. The estimated irradiance in the absence of plasma was $I_{L,Vac}=P_L/\pi r_{Eff}^2$=(7.9±1.4)×10$^{18}$ W/cm$^2$, and the corresponding dimensionless amplitude was $a_{0,Vac}$=1.9±0.2. In helium plasma with the initial electron density $n_e$≈(2.6±0.8)×10$^{19}$ cm$^{-3}$, these values increased significantly due to relativistic self-focusing [24] up to an estimated value $a_{SF}$≈7 assuming stationary self-focusing [25]. Moreover, after relativistic self-focusing, the condition of bow wave generation [3] is always satisfied, thus enabling excitation of prominent bow waves, which in turn produced singularities of relativistic multistream plasma flows emitting BISER [2]. We used a carefully calibrated spectrograph, similar to the one described in [5]. However, this time we used Al optical blocking filters, which provided the 17–34 nm spectral range defined by the Al L$_{2,3}$ absorption edge at ~17 nm [26] and its second spectral order.

A typical spectrum with fine fringes is shown in Fig. 2(ab); its Fourier transform (i.e. Fourier transform of its *intensity*, not including the spectral phase which was not measured), Fig. 3(a) black line, contains several components, including a prominent feature at ~7.6 eV$^{-1}$; this feature is dominant in a local Fourier transform with a 38±2 eV Gaussian window, red line.

We performed Fourier analysis of 482 shots and extracted spectral fringe separations in two spectral regions, 36-40 eV and 41-48 eV with corresponding Nyquist frequencies of ~19 eV$^{-1}$ and ~15 eV$^{-1}$ (determined by two times the separations between the x-ray CCD pixels). Advantages of these two spectral regions include absence of artifacts from He plasma emission lines and higher spectrograph resolving power, compared to the 55-72 eV spectral region. Two fringe spacings were extracted in each case, one with the dominant peak in the Fourier space (the frequency of the strongest modulation), and another with the lowest modulation frequency (i.e. the closest to the laser frequency). The results are shown in Fig. 3(b). The distributions have several peaks, most prominently at ~1 and 2 eV$^{-1}$, and less prominently around 2.9 eV$^{-1}$, with ratios approximately 1:2:3; there are further peaks at higher fringe frequencies corresponding to even finer fringes.

pulse propagating through plasma experiences frequency downshift and dispersion, and by the time BISER is generated its photon energy becomes $\hbar\omega_0$~0.5–1 eV, as suggested by the first peak with fringe frequencies ~1–2 eV$^{-1}$ in the distribution, Fig. 2(b).

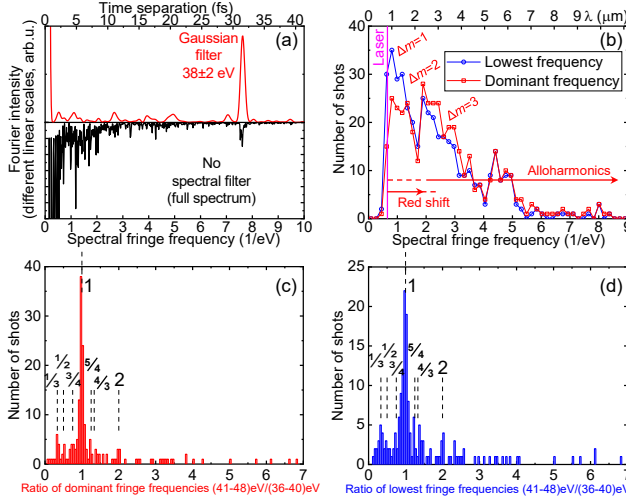

FIG. 3. Analysis of experimental BISER spectra. (a) Fourier transforms of the full spectrum shown in Fig. 2(a) (black) and its portion limited by a 38±2 eV Gaussian window (red). (b) Distribution of dominant and lowest fringe frequencies in the 36-40 eV and 41-49 eV spectral ranges. (c,d) Histograms of ratios of fringe frequencies in the 41-48 eV and 36-40 eV spectral ranges. (b)-(d) contain data from 482 laser shots; spectral fringes have been observed in 195 and 234 shots in the 36-40 eV and 41-48 eV spectral regions, respectively.

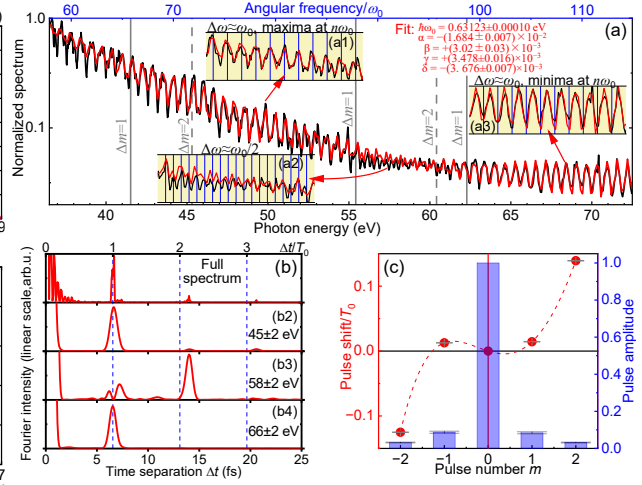

FIG. 4. Analysis of the spectrum shown in Fig. 2(c). (a) Experimental spectrum (black) and its alloharmonic fit (red, see text) [20]. Insets (a$_1$-a$_3$) show regions with fringe spacing $\Delta\omega \approx \omega_0$, $\approx \omega_0/2$, and $\approx \omega_0$, respectively; further, fringes in (a$_1$) and (a$_3$) exhibit counter-phase positions relative to the $n\omega_0$ grid shown by thin vertical lines. The gray vertical lines show alloharmonics with $\Delta m=1$ (solid) and $\Delta m=2$ (dashed) determining the main fringe periodicities. (b) Fourier transforms of the full spectrum (top) and its windowed regions. (c) Properties of the aperiodic pulse train (amplitudes and shifts from the periodic positions) providing the best fit.

In some shots, different fringe periodicities were observed in the 36-40 and 41-48 eV spectral ranges. For each shot, we calculated the ratio of fringe frequencies from these spectral ranges, and plotted histograms of these ratios, Figs. 3(c,d). Along with the peaks at ratio ≈1 (i.e. nearly the same fringe spacing), there were lower peaks with *approximately k:l* ratios with small integers $k$ and $l$, indicating a non-random ratio of fringe periods in these two spectral ranges in each single shot.

Moreover, we observed several fringe periods in many single shots. An example is shown in Fig. 2(c), its spectrum in Fig. 4(a) by the black line, and Fourier analysis of the full spectrum and its windowed parts in Fig. 4(b). Again, the ratio of the fringe spacings of approximately 1:2:3 is obvious.

The observed features can be explained by the combined effect of the gradual frequency downshift [14,15] and alloharmonics [20]. Namely, the laser

The laser pulse, as it generates BISER, continues to get redshifted, and propagates through gradually changing plasma density. Thus, BISER is emitted as a nonperiodic train containing a few attosecond pulses. The fringes observed experimentally are alloharmonics with spacing $\Delta\hbar\omega \approx \hbar\omega_0/\Delta m$ with $\omega_0$ corresponding to the redshifted laser frequency at the BISER emission time; thus, the $\hbar\omega_0$ distribution is continuous within the ~0.5−1 eV range corresponding to shot-to-shot redshift and BISER emission time variations. On the other hand, discrete (integer) values $\Delta m=m_2-m_1$ correspond to different pairs of cycle numbers $m_1$ and $m_2$ contributing to the spectral interference in the particular spectral range; the maximum $\Delta m$ value is determined by the number of pulses in the train. Alloharmonics explain peaks in Fig. 2(b) with a ratio of approximately 1:2:3 as $\Delta m$ changes across the shots, small peaks with ratios 2:1, 1:2, 1:3,… in Figs. 3(c,d), and ≈1:2:3 ratios in Fig. 4(b) as

Δm changes in the same shot, depending on the spectral range.

Further, alloharmonics explained the complex spectral shapes of the observed BISER spectra, exemplified by the black curve in Fig. 4(a). We fitted this experimental spectrum with an alloharmonic model with the frequency vs time dependence $\omega(t)=\omega_0(1+2\alpha(t/T_0)+3\beta(t/T_0)^2+4\gamma(t/T_0)^3+5\delta(t/T_0)^4)$, and found excellent agreement, red curve in Fig. 4(a). Here the best-fit central frequency, $\hbar\omega_0=0.63123\pm0.00010$ eV, is significantly redshifted compared to the laser photon energy, $\hbar\omega_L=1.5278$ eV. The pulse train parameters, i.e. pulse amplitudes and shifts from periodic positions, are shown in Fig. 4(c). This fit is discussed in more detail in [20].

Alloharmonics also explained significant shot-to-shot variability of the observed spectra. Indeed, change in the frequency variation rate (the chirp parameters) by a few % significantly changes the high-frequency part of the emitted spectrum [9]. This property allows precise driver central frequency and nonperiodicity parameters determination via fitting of the observed high-frequency spectrum; an example of this procedure applied to our data is shown in Fig. 4(a).

Resolving alloharmonics with frequency separations several or many times smaller than the laser frequency becomes increasingly challenging with increasing photon energy. This is relevant not only to BISER, but to all "harmonic generation" mechanisms where the driver is at least slightly nonperiodic, which is practically always the case. In the present experiment, around half of the shots exhibited no spectral fringes, while in the earlier experiments [4,5], even a larger portion of the BISER spectra contained no fine fringes, probably due to insufficient spectral resolving power at higher photon energies. In atomic and Sliding Mirror [27] harmonics, such absence of fringes is used as an important signature of an Isolated Attosecond Pulse (IAP) generation [28]. In experiments with other x-ray sources, the absence of spectral fringes may indicate an absence of temporal coherence. However, our present analysis showed that the absence of fringes at high frequencies can be due to the insufficient spectrograph resolving power, so the conclusions about the fringe absence reasons should be derived with care.

As the general nature of alloharmonics suggest, they are present not only in BISER spectra and in electromagnetic waves. An example of alloharmonics in a gravitational wave is shown in Fig. 5. We took advantage of the strain vs time dependences calculated by a numerical relativity technique in the literature [29], a binary blackhole case SXS:BBH:1148, with a mass ratio of 2.038, dimensionless spins of 0.43 and 0.51, and initial orbital angular frequency of $\omega_{Orb,ini}=9.165\times10^{-3}/M$, Fig. 5(a). Although gravitational spectrometers are not available yet, these published dependences allowed us to calculate spectra, Fig. 5(b), as may be recorded by future gravitational spectrometers with a millisecond-scale time window. We used window durations of $1000M$ (geometric units), where $M$ is the sum of gravitating masses; for 20 solar masses, this window duration corresponds to 100 ms. The gravitational wave spectrum initially contains harmonics, however, as the inspiral progresses, alloharmonics appear at frequencies 3-4 times larger than the instantaneous inspiral frequency, dashed vertical lines in Fig. 5(b). Importantly, the value of the dimensionless chirp parameter, $\alpha_{GW}=0.0175$, in the blue window in Fig. 5(a), is very similar to the absolute value derived from the BISER experiment, $\alpha_{BISER}=-(1.684\pm0.007)\times10^{-2}$, which provides the basis for scaled laboratory astrophysics [30,31] experiments, as the chirp sign corresponds to time flipping and does not affect the spectral shape.

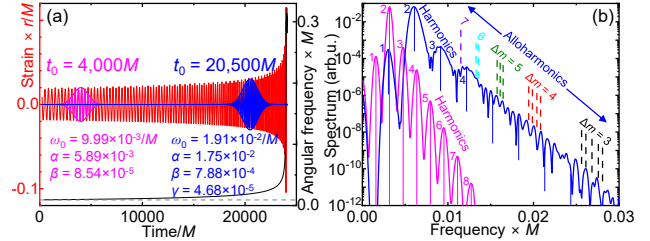

FIG. 5. Alloharmonics in a gravitational wave emitted by an inspiralling black hole binary [20]. (a) Strain vs time (red), numerical relativity calculation taken from the database [29]; pink and blue lines show two different time windows of interest. The dashed line shows the initial inspiralling frequency, the black line shows angular frequency vs time, from which the central angular frequency $\omega_0$ and dimensionless chirp parameters $\alpha$, $\beta$, and $\gamma$ are calculated for both windows. (b) Gravitational wave spectra of time windows shown in (a) by pink and blue lines, respectively. The vertical solid and dashed lines show calculated positions of harmonics and alloharmonics, respectively. We use geometric units: the time unit is $GM/c^3$, the unit of distance $r$ is $GM/c^2$, where $G$ is the gravitational constant. For the sum of gravitating masses $M=M_\odot$ (one solar mass), these are 4.93 μs and 1.48 km, respectively.

Other examples of alloharmonics in slightly nonperiodic optical frequency combs and radio emission of slowing-down pulsars are provided in [20].

In conclusion, we found that the observed fine BISER spectral fringes with various separations up to ~13 times smaller (0.12 eV) than the initial laser frequency (~1.5 eV) are explained by the combination of two effects: the gradual laser frequency downshift during propagation through plasma (factors of 1.5-3), and alloharmonics produced by driver nonperiodicity (additional close-to-integer factors ~1, ~2, ~3, etc.). Alloharmonics may have a much smaller fringe separation than harmonics, which must be considered in designing experiments on spectral and attosecond temporal x-ray pulse characterization, irrespective of the coherent x-ray generation mechanism. Resolving these fringes becomes especially challenging at high frequencies, such as in the keV spectral region [32]. Alloharmonics is a general phenomenon present not only in BISER spectra but in all electromagnetic, gravitational, acoustic, etc. waves containing harmonics generated by an even slightly nonperiodic driver.

## ACKNOWLEDGMENTS

We are grateful to the J-KAREN-P laser team and acknowledge discussions with Dr. S. B. Echmaev. This research was supported by JSPS Kakenhi JP26707031, JP19H00669, JP19KK0355, JP23H01151, and JP25H00621 and Strategic Grants by the QST President: Creative Research #20 and IRI.